\documentclass[aps,prd,preprint,superscriptaddress]{revtex4}
\usepackage{graphicx}
\usepackage{amsfonts,amsmath,amssymb,bm}
\begin{document}

\markboth{Peralta-Ramos, J., Nakwacki, M.S.}{Production of thermal photons in viscous fluid dynamics with temperature-dependent shear viscosity}


\title{Production of thermal photons in viscous fluid dynamics with temperature-dependent shear viscosity}

\author{J. Peralta-Ramos}

\address{Instituto de F\'isica Te\'orica, Universidade Estadual Paulista \\
Rua Doutor Bento Teobaldo Ferraz, 271 - Bloco II, S\~ao Paulo, SP, Brazil\\
jperalta@ift.unesp.br}

\author{M.S. Nakwacki}

\address{Instituto de Astronom\'ia, Geof\'isica e Ci\^encias Atmosf\'ericas, Universidade de S\~ao Paulo, \\ 
Rua do Mat\~ao 1226, Cidade Universit\'aria, 05508-090 S\~ao Paulo, Brazil\\
sole@astro.iag.usp.br}


\begin{abstract}
We compute the spectrum of thermal photons created in Au+Au collisions at $\sqrt{s_{NN}}=200$ GeV, taking into account dissipative corrections in production processes corresponding to the quark--gluon plasma and hadronic phases. To describe the evolution of the fireball we use a viscous fluid dynamic model with different parametrizations for the temperature--dependence of $\eta/s$. We find that the spectrum significantly depends on the values of $\eta/s$ in the QGP phase, and is almost insensitive to the values in the hadronic phase. 
We also compare the influence of the  temperature--dependence of $\eta/s$ on the spectrum of thermal photons to that of using different equations of state in the fluid dynamic simulations, finding that both effects are of the same order of magnitude.
\end{abstract}
\maketitle

\section{Introduction}

There is currently great interest and effort in understanding the transport properties of the quark-gluon plasma (QGP) and of the hadronic matter created in ultrarelativistic heavy ion collisions at RHIC and LHC \cite{latest,revhydro,heinz}. As evidenced by the large elliptic flow that is measured, matter created in these events behaves as a nearly perfect fluid with a very low viscosity--to--entropy ratio $\eta/s$. 
For this reason, relativistic viscous fluid dynamics with $\eta/s \lesssim 0.4$ has been very successful in describing the hadronic observables such as the radial and elliptic flows \cite{latest,h1,song,h1bis,h2,h3,h4,h4bis,h5,est}. For recent reviews on the application of relativistic hydrodynamics to heavy ion collisions see e.g. \cite{revhydro,heinz}. 
   
Great efforts are currently focused on developing new theoretical tools to compute more accurately the transport coefficients of the QGP and hadronic matter from microscopic models (see e.g. \cite{tr1,tr2,tr3,brsss,gravity,natsuume}) as well as on extracting them more precisely from RHIC and LHC measurements \cite{revhydro,heinz,h1,song,h2,h1bis,h3,h4,h4bis,h5,est}.  Although it is expected that the transport coefficients of the QGP depend on temperature, the impact of a temperature--dependent $\eta/s$ on momentum anisotropies as obtained from simulations has been investigated only recently \cite{h1bis,h2,Nagle:2011uz,bozek,koide,songheinz,jpr-krein}. These studies show that the momentum anisotropies are indeed affected by the tempereature dependence of $\eta/s$, pointing to the conclusion that further investigation is necessary to be able to extract precise values of $\eta/s$ from data.

In this paper we focus on the spectrum of thermal photons produced during the expansion of the fireball created in Au+Au collisions at $\sqrt{s_{NN}}=200$ GeV. In constrast to hadrons, the photons created in the interior of the fireball pass through it without any interaction, thus giving information on the properties of {\it bulk} nuclear matter rather than on its surface \cite{phenix,rev,nos,bhat,bhat2,dus,gale,gale2,bleich,bleich2,bleich3,elli1,elli2,mitra}. 

The main purpose of this work is to quantify the sensitivity of the spectrum of thermal photons to the temperature--dependence of $\eta/s$. To this end, we compute the spectrum of thermal photons as obtained from fluid dynamic simulations using simple parametrizations of $\eta/s$ as a function of temperature, which allow us to disentangle the impact that the temperature dependence of $\eta/s$ in the different phases (QGP and hadronic) has on thermal photon spectra.  

Thermal photon spectra have been calculated within the framework of ideal hydrodynamics (see \cite{rev}), using the Israel--Stewart formalism \cite{bhat,bhat2,dus} and more recently using a viscous $(3+1)$D implementation \cite{elli1}. The inclusion of viscous corrections during the QGP phase was studied in \cite{bhat,bhat2} using Israel--Stewart theory and taking into account both shear and bulk viscosity, while in \cite{dus} the nonequilibrium correction to photon production due to Compton scattering and $q\bar{q}$ annihilation were computed at leading--log order. Recently, thermal photon spectra including dissipative corrections has been computed in a $(1+1)$D hydrodynamic model including shear and bulk viscosities and with an EOS corresponding to a first order phase transition \cite{mitra}. The spectrum of thermal photons has also been computed using hydro-kinetic hybrid approaches to describe the evolution of the fireball \cite{bleich,bleich2,bleich3}. 
In \cite{nos} we compared the thermal photon spectra obtained from second order fluid dynamics and a divergence-type theory that includes dissipative effects to all orders in velocity gradients \cite{est}, concluding that differences in dissipative fluid dynamic models are a significant source of uncertainty in the precise determination of $\eta/s$ from data. It is important to notice that there has been an increasing interest in the elliptic flow of photons produced in heavy ion collisions, since this observable may provide deeper understanding on the properties of the strongly coupled QGP -- see \cite{elli1,elli2} where the elliptic flow and the spectrum of photons are calculated.

It is appropriate at this point to comment on the limitated scope of our work. On the fluid dynamics side, we neglect bulk viscosity, which peaks at the critical temperature and is known to increase photon production \cite{bhat,bhat2}. Besides, our model is $(2+1)$D boost invariant. This approximation is not so limiting if one is concerned only with the dynamics
near the mid--rapidity region, which is the case studied here. We also assume that the baryonic chemical potential is
zero in the central rapidity region, and therefore the conservation equation of the net baryon number is not considered. 
Moreover, we use smooth initial conditions for the energy density. The importance of initial--state fluctuations and its impact on the hydrodynamic flow and final observables has only recently been noted (see \cite{lump1,lump2,lump3}). In particular, as shown in \cite{chatt,chatt2}, event--by--event fluctuations of the initial energy density produced in
heavy ion collisions at RHIC significantly enhance the production of thermal photons as compared to a smooth initial--state averaged profile in ideal hydrodynamic simulations. 
On the photon production side, we do not take into account prompt and jet--medium photons, whose contribution to the total photon spectrum becomes important at large transverse momentum \cite{gale,gale2}. In spite of these limitations, the model employed here describes realistically the hydrodynamic evolution of matter created at RHIC (see \cite{h3,est,jpr-krein}), allowing us to quantify the impact of the temperature--dependence of $\eta/s$ on the spectrum of thermal photons. 

This paper is organized as follows. In Section \ref{model} we describe the viscous fluid dynamic model and provide numerical details of the simulations, and also describe the procedure used to compute the photon spectra. In Section \ref{res} we present and discuss our results, and finally we conclude in Section \ref{conc}. 

\section{Theoretical model}
\label{model}

In this section, we first give a brief overview of the equations of viscous fluid dynamics used to model the evolution of matter created in heavy ion collisions, and provide details of the simulations performed. Then we go over to describe the processes that taken into account to compute the thermal photon spectra.

\subsection{Viscous fluid dynamics}
\label{hydro}

We will now briefly review the $(2+1)$D hydrodynamic model used in this paper. A detailed description of the model as applied to heavy ion collisions can be found in \cite{h3,est}. 

In what follows, Latin indices stand for transverse coordinates $(x,y)$, $D^\mu$ is the geometric covariant derivative, $D=u_\mu D^\mu$ and $\nabla^\mu=\Delta^{\mu\nu}D_\nu$ are the comoving time and space derivatives, respectively, and brackets around indices  imply taking the spatial, symmetric and traceless projection of a tensor. We employ Milne coordinates defined by proper time $\tau=\sqrt{t^2-z^2}$ and rapidity $\psi=\textrm{arctanh}(z/t)$, and work in flat space-time. We assume boost invariance so that all quantities are independent of $\psi$. The fluid velocity is $\vec{u}=(u^\tau,u^x,u^y,0)$ and is normalized as $u_\mu u^\mu=1$. 
The stress-energy tensor reads $T^{\mu\nu} =\rho u^\mu u^\nu - p \Delta^{\mu\nu} + \Pi^{\mu\nu}$, with 
$\Delta^{\mu\nu} = g^{\mu\nu}-u^\mu u^\nu$, where $\rho$ and $p$ are the energy density and the pressure in the local rest frame, and $\Pi^{\mu\nu}$ is the viscous shear tensor. For a conformal fluid as considered here, $T^\mu_\mu=0$, so $\rho=3p$ and the bulk viscosity vanishes. 

The hydrodynamic equations are the conservation equations for the stress-energy tensor together with the evolution equation for the shear tensor $\Pi^{\mu\nu}$. The latter reads \cite{brsss,gravity}
\begin{equation}
\begin{split}
\partial_\tau \Pi^{i\alpha} &= -\frac{4}{3u^\tau}\Pi^{i\alpha}\nabla_\mu u^\mu - \frac{1}{\tau_\pi u^\tau}\Pi^{i\alpha} + \frac{\eta}{\tau_\pi u^\tau} \sigma^{i\alpha} \\
&~ - \frac{\lambda_1}{2\tau_\pi \eta^2 u^\tau}\Pi^{<i}_\mu \Pi^{\alpha> \mu} - \frac{u^i\Pi^\alpha_\mu + u^\alpha \Pi^i_\mu}{u^\tau}Du^\mu \\
&~ -\frac{u^j}{u^\tau}\partial_j \Pi^{i\alpha} 
\end{split}
\label{dpi}
\end{equation}
where $\eta$ is the shear viscosity, $(\tau_\pi,\lambda_1)$ are second-order transport coefficients, and $\sigma^{\mu\nu}=\nabla^{<\mu}u^{\nu>}$ is the first-order shear tensor.

The conservation equations for $T^{\mu\nu}$ and the evolution equation for $\Pi^{\mu\nu}$ given in Eq. (\ref{dpi}) are solved selfconsistently using the parameters defined in the following section. 


\subsection{Numerical setup}
\label{num}

We choose $(\rho,u^x,u^y,\Pi^{xx},\Pi^{xy},\Pi^{yy})$ as independent variables. Solution of the hydrodynamic equations requires initial conditions for the six independent variables, which we take to be $u^x=u^y=0$ and $(\Pi^{xx},\Pi^{xy},\Pi^{yy})=0$, while the initial energy density profile is calculated using a simple Glauber model \cite{est}. 
As in previous fluid dynamic simulations of heavy ion collisions at $\sqrt{s_{NN}}=200$ GeV performed with the same model \cite{est}, we take the initialization time to be $\tau_0=$ 1 fm/c, use a computational grid of 13$\times$13 fm and 
use values for the second-order transport coefficients corresponding to a strongly-coupled $\cal{N}=$ 4 Super-Yang Mills (SYM) plasma \cite{h3,est,brsss,gravity,natsuume}, $\tau_\pi = 2(2-\ln 2) \eta/(sT)$ and $\lambda_1=\eta/(2\pi T)$, where $s$ is the entropy density and $T$ is the temperature.

\begin{figure}[pb]
\scalebox{0.8}{\includegraphics{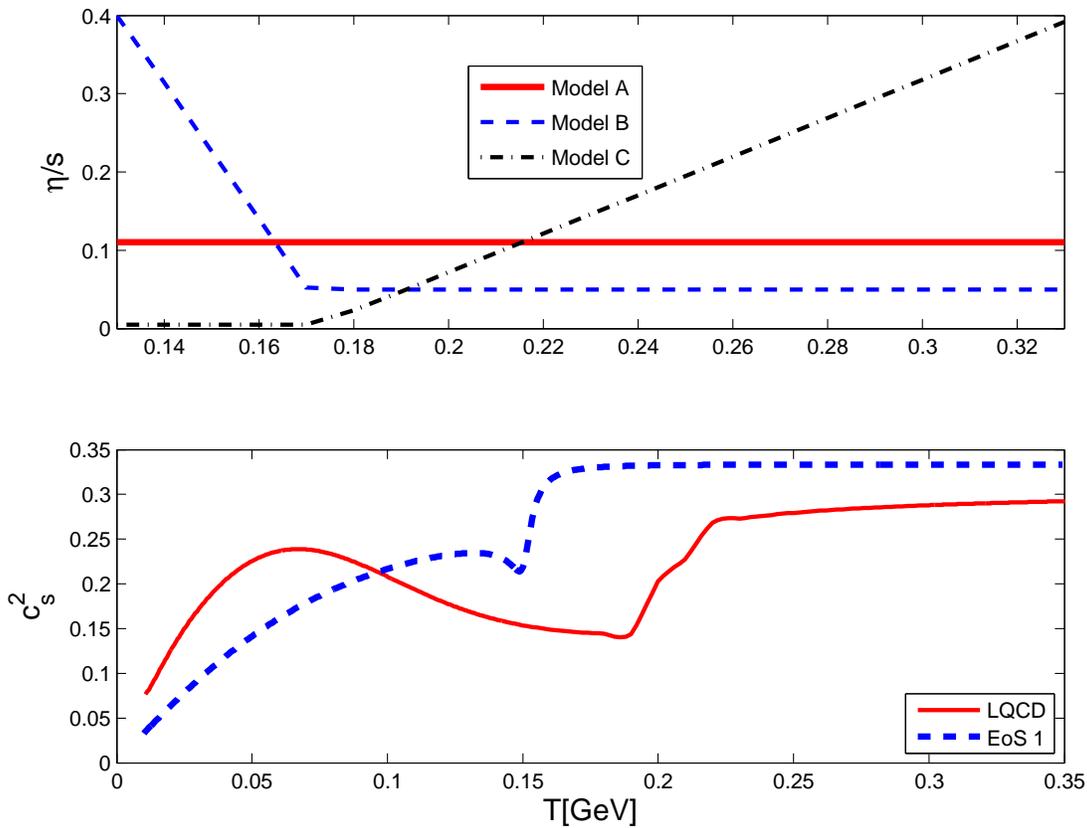}}
\vspace*{8pt}
\caption{(Color online) Three different parametrizations for the temperature dependence of $\eta/s$: constant (Model A), constant in the QGP phase and varying in the hadronic phase (Model B), and constant in the hadronic phase and varying in the QGP phase (Model C) ({\it upper panel}). Equation of state used in the simulations (denoted by LQCD) that was obtained by Laine and Schr\"{o}der$^{43}$, together with a different EOS (denoted by EOS 1) that is used for comparison ({\it lower panel}).}
\label{eta}
\end{figure}

To solve the hydrodynamic equations it is necessary to provide the value of $\eta/s$ as an input, as well as an equation of state (EOS) relating the pressure and energy density. Here we will employ three parametrizations of $\eta/s$ as a function of temperature.  Specifically, we consider the cases in which the ratio stays constant throughout the entire fluid dynamic evolution (Model A), or varies in the hadronic phase and stays constant in the QGP phase (Model B), or else varies in the QGP phase and stays constant in the hadronic phase (Model C). The three cases are shown in the upper panel of Figure \ref{eta}. We note that for Model A we choose the value of $\eta/s$ to be equal to the average value obtained from Models B and C, namely $\eta/s=0.11$, which gives sense to the comparison between the three parametrizations. This value of $\eta/s$ is on the lower side in terms of fitting charged hadron elliptic flow to data \cite{h3,est}. 

With respect to the EOS, we employ the EOS obtained by Laine and Schr\"{o}der \cite{laine}, which connects a high-order weak-coupling perturbative QCD calculation at high temperatures to a hadron resonance gas at low temperatures, via an analytic crossover as suggested by Lattice QCD calculations -- see e.g. \cite{aoki,huov,huov2,huov3,hirano,heinz05}. We notice that this EOS is the same as that used in previous simulations focusing on charged hadron elliptic flow \cite{h3,est}. The EOS is shown in the lower panel of Figure \ref{eta}, together with a different EOS \cite{jpr-krein} that is used later on for comparison (see Section \ref{res}).

Since the model used here is purely hydrodynamic, a cut--off for the value of $\eta/s$ must be 
imposed in order to avoid the breakdown of the fluid description. The value of $\eta/s$ at which it is sensible to impose this cut--off is 
constrained by the comparison of results obtained from hydrodynamic and kinetic simulations to 
data. In our simulations we set $\eta/s \leq 0.4$, which is close to the upper bound imposed by such comparisons to data -- see \cite{jpr-krein}. 
We note that the conclusions extracted from our results do not depend on the precise value of the cut--off for the value $\eta/s$.

\subsection{Thermal photon production}
\label{phot}

In order to compute the spectrum of thermal photons created during the evolution of the fireball we consider the processes of Compton scattering, $q\bar{q}$ annihilation and bremsstrahlung in the QGP phase, and $\pi\pi \rightarrow \rho \gamma$, $\pi\rho \rightarrow \pi \gamma$ and $\rho \rightarrow \pi \pi \gamma$ in the hadron phase \cite{rev,nos}. In all cases, the 
nonequilibrium distribution function of the quarks, $f(x^\mu,p^\mu)$, is calculated from Grad's quadratic ansatz \cite{revhydro,heinz,h3,est}:
\begin{equation}
f(x^\mu,p^\mu) = f_0+f_0(1-f_0)\frac{p^\mu p^\nu\Pi_{\mu\nu}}{2T^2 (\rho + p)}
\label{dis}
\end{equation}
where $f_0$ is the equilibrium distribution function of the quarks. The evolution of the temperature $T$, energy density $\rho$, fluid velocity $u^\mu$ and shear tensor $\Pi^{\mu\nu}$ are obtained from the fluid dynamic model described in the previous section. 

The production rate for Compton scattering and $q\bar{q}$ annihilation reads \cite{rev,dus}
\begin{equation}
E\frac{dN}{d^4xd^3p}=\frac{1}{2\pi^2}\alpha\alpha_s(\sum_fe^2_f)T^2 f(x^\mu,p^\mu) \ln\bigg(\frac{cE}{\alpha_sT}\bigg)
\end{equation}
with $c\sim$ 0.23, $\alpha=1/137$ and \cite{as} $\alpha_s(T) =6\pi/((33-2N_f)\ln(8T/T_c))$, where $N_f$ is the number of quark flavors, $e_f$ is the electric charge of the quark and $T_c$ is the critical temperature, which we set to be $T_c=170$ MeV in agreement with Lattice QCD simulations \cite{aoki}. The contribution due to $q\bar{q}$ annihilation when an additional scattering is included is 
\begin{equation}
E\frac{dN}{d^4xd^3p}=\frac{8}{3\pi^5}\alpha\alpha_s(\sum_fe^2_f)ETf(x^\mu,p^\mu)[J_T-J_L] ~,
\end{equation}
with $J_T\sim$ 1.11 and $J_L \sim$ 1.06. 

For bremsstrahlung we have \cite{brem}
\begin{eqnarray}
E\frac{dN}{d^4xd^3p}&=&\frac{8}{\pi^5}\alpha\alpha_s(\sum_fe^2_f)\frac{T^4}{E^2}f(x^\mu,p^\mu)(J_T-J_L)\times \nonumber\\
& &\bigg[3\zeta(3) +\frac{\pi^2E}{6T} +\frac{E^2}{T^2}\ln 2 +4\textrm{Li}_3(-f(x^\mu,p^\mu))\nonumber\\
&+&2\frac{E}{T}\textrm{Li}_2(-f(x^\mu,p^\mu))-\frac{E^2}{T^2}
\ln(1+f(x^\mu,p^\mu))\bigg]
\end{eqnarray}
where $\zeta$ is the zeta function and Li$_m=\sum_{n=1}^\infty z^n/n^m$ are polylog functions. 

For the production rate of photons during the hadronic phase we use the estimate calculated by Steffen and Thoma \cite{hhg} (see also \cite{nad}), which reproduces the sum of production rates for the processes $\pi\pi \rightarrow \rho \gamma$, $\pi\rho \rightarrow \pi \gamma$ and $\rho \rightarrow \pi \pi \gamma$:
\begin{equation}
E\frac{dN}{d^4xd^3p} \simeq ~4.8 T^{2.15} e^{-1/(1.35 ET)^{0.77}} f(x^\mu,p^\mu) ~.
\end{equation}
 
The total thermal photon spectrum is then obtained by integrating the sum of the production rates over the evolution of the fireball as obtained from fluid dynamics, thus 
\begin{equation}
 \bigg(\frac{dN}{d^2p_TdY}\bigg)=\int d\vec{x}\int^{\tau_2}_{\tau_1}\int^{Y}_{-Y}E\frac{dN}{d^4xd^3p}
\end{equation}
where $\tau_{1,2}$ are the initial and final times of each phase, $Y$ is the rapidity of the nuclei, $d\vec{x}=(dx,dy)$ and $p_T$ is the transverse momentum. Note that the photon energy in the comoving frame is given by $p_T \textrm{cosh}(Y-Y')$. In what follows we will limit ourselves to the case $Y=Y'=0$.

\section{Results}
\label{res}

In this section we present and analyze the spectra of thermal photons computed from the fluid dynamic model described above. 

\begin{figure}[pb]
\scalebox{0.8}{\includegraphics{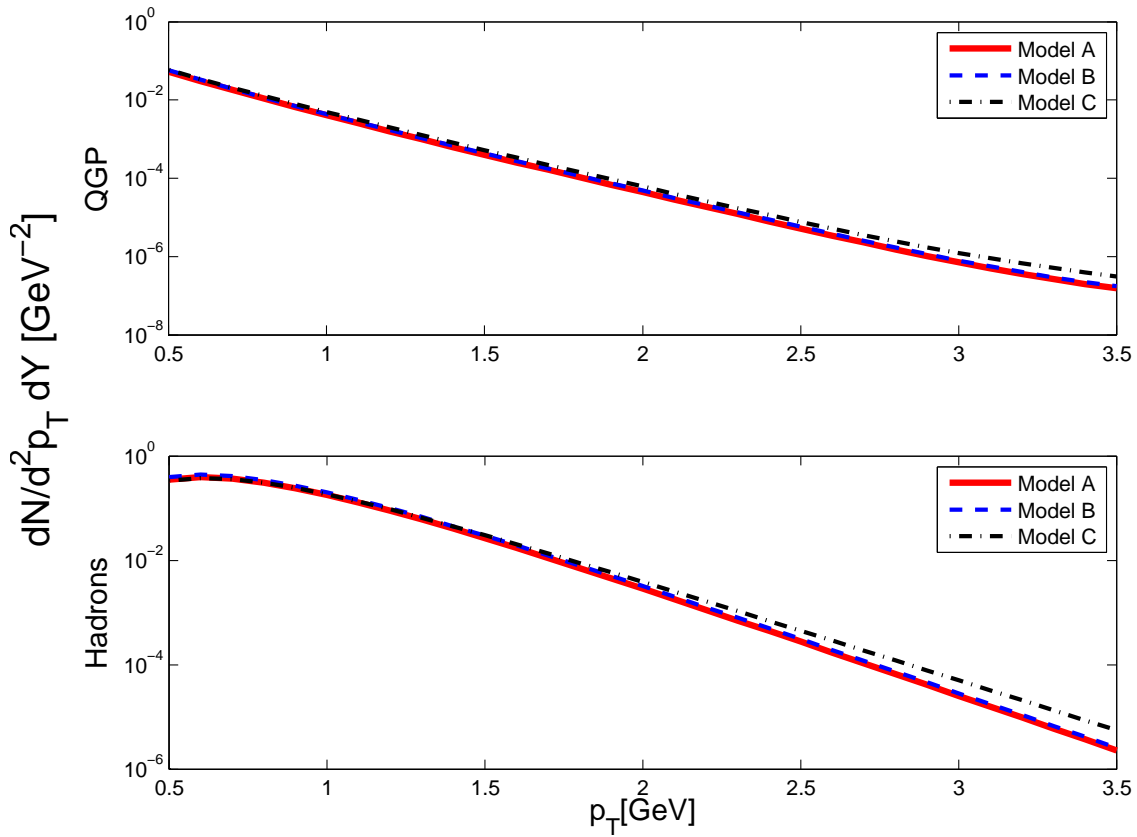}}
\vspace*{8pt}
\caption{(Color online) Contributions to the thermal photon spectrum corresponding to the QGP ({\it upper panel}) and the hadronic phases ({\it lower panel}), for Models A (temperature--independent), B (varying in the hadronic phase) and C (varying in the QGP phase) for $\eta/s$.}
\label{qgp-h}
\end{figure}

Figure \ref{qgp-h} shows the contributions to the thermal photon spectrum corresponding to the QGP ({\it upper panel}) and the hadronic phases ({\it lower panel}), for Models A (temperature--independent), B (varying in the hadronic phase) and C (varying in the QGP phase) for $\eta/s$. It is seen that the difference in the spectra corresponding to Models A and B is very small in the whole range of values of $p_T$ that we consider.  In contrast, the spectrum computed from Model C is significantly larger, and the difference with the spectra of Models A and B increases with increasing values of $p_T$. It is also seen that the difference between the results obtained from Model C and those obtained from Models A and B is slightly larger for the hadronic phase. 

Previous studies focusing on the elliptic flow of charged hadrons computed from viscous fluid dynamics with different profiles of $\eta/s$ as functions of temperature \cite{h1bis,h2,Nagle:2011uz,bozek,koide,songheinz}, showed that at RHIC energies the elliptic flow is almost insensitive to the value of $\eta/s$ in the QGP phase but depends strongly on the value of $\eta/s$ in the hadronic phase. Our results show that, at least under the approximations made here, for the thermal photon spectrum the situation is quite the opposite: the spectrum is much more sensitive to the QGP shear viscosity than to the values in the hadronic phase. As shown in Figure \ref{qgp-h}, the temperature dependence of the ratio $\eta/s$ in the hadronic phase (which corresponds to Model B) leads to a thermal photon spectrum that is almost indistinguishable from the spectrum computed from a temperature--independent value of $\eta/s$ (Model A), in sharp contrast to what happens with the spectrum obtained from an $\eta/s$ that varies in the QGP phase (which corresponds to Model C). 
One should be careful to note that small $p_T$ photons are mostly produced during the hadronic phase, whereas large $p_T$ photons are mainly produced in the QGP phase. Therefore, it is perhaps more accurate to conclude that large $p_T$ photons are more sensitive than small $p_T$ photons to variations in $\eta/s$ . This result implies that serious difficulties would arise when attempting to extract the temperature--dependence of $\eta/s$ corresponding to the hadronic phase by matching fluid dynamic simulations to measured photon spectra. 

Although it is tempting to conclude that fitting the spectrum of thermal photons to fluid dynamic simulations based on different parametrizations of temperature--dependent $\eta/s$ would allows us to directly extract this temperature dependence in the QGP phase from data, in reality much more work is needed to reach this stage, essentially because it is still necessary to determine more precisely the uncertainty coming from various sources in viscous fluid dynamic simulations. An example of such a source of uncertainty in the possible extraction of $\eta/s$ from the measured photon spectrum is presented later on when discussing results obtained with different EOSs.

\begin{figure}[pb]
\scalebox{0.8}{\includegraphics{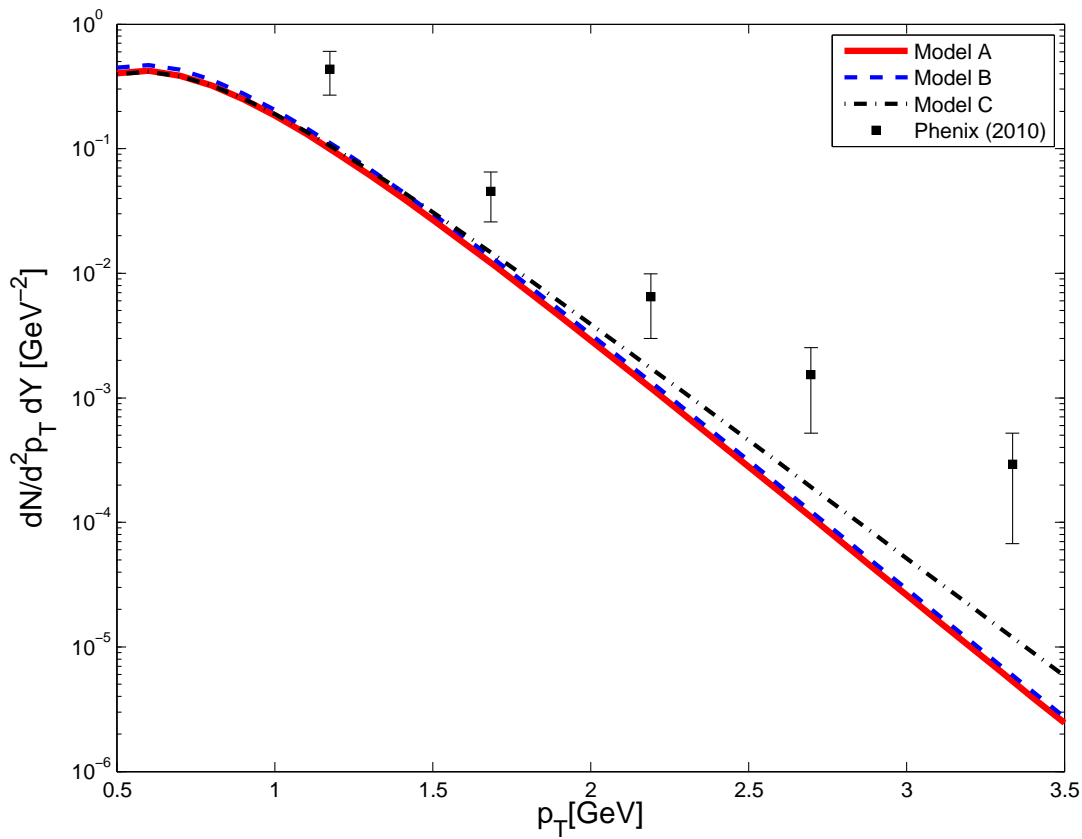}}
\vspace*{8pt}
\caption{(Color online) Total thermal photon spectrum corresponding to Models A (temperature--independent), B (varying in the hadronic phase) and C (varying in the QGP phase) for $\eta/s$. The data corresponds to direct photons as measured by the Phenix Collaboration$^{24}$.}
\label{total}
\end{figure}

Returning to our results, Figure \ref{total} shows the total thermal photon spectrum corresponding to the three models for $\eta/s$. Also shown are data of direct photons as measured by the Phenix Collaboration \cite{phenix}. We note that we do not attempt to fit our results to data, since as mentioned in the Introduction we neglect prompt and jet--medium photons in calculating the total spectrum, considering only thermal photons. The rationale for showing experimental results is merely to show that our viscous fluid dynamic model is not excluded by data, which would definitively be the case if the computed spectrum were larger that the observed one.
It is seen from Figure \ref{total} that, as expected from the results shown in Figure \ref{qgp-h}, the total spectrum corresponding to Model C is considerably larger than those corresponding to Models A and B. Moreover, the difference between the results obtained from Model C and Models A-B  increases with increasing values of $p_T$ and becomes significant for $p_T \geq 1.5$ GeV.

To put our results into perspective, it is useful to compare the impact that the temperature--dependence of $\eta/s$ has on thermal photon spectra to the effect of another input of fluid dynamic simulations, namely the EOS. This allows us to determine the relative importance of the temperature--dependence of $\eta/s$ on the spectra as compared to another important source of uncertainty in fluid dynamic simulations as is the precise shape of the EOS of nuclear matter created in heavy ion collisions. 

\begin{figure}[pb]
\scalebox{0.8}{\includegraphics{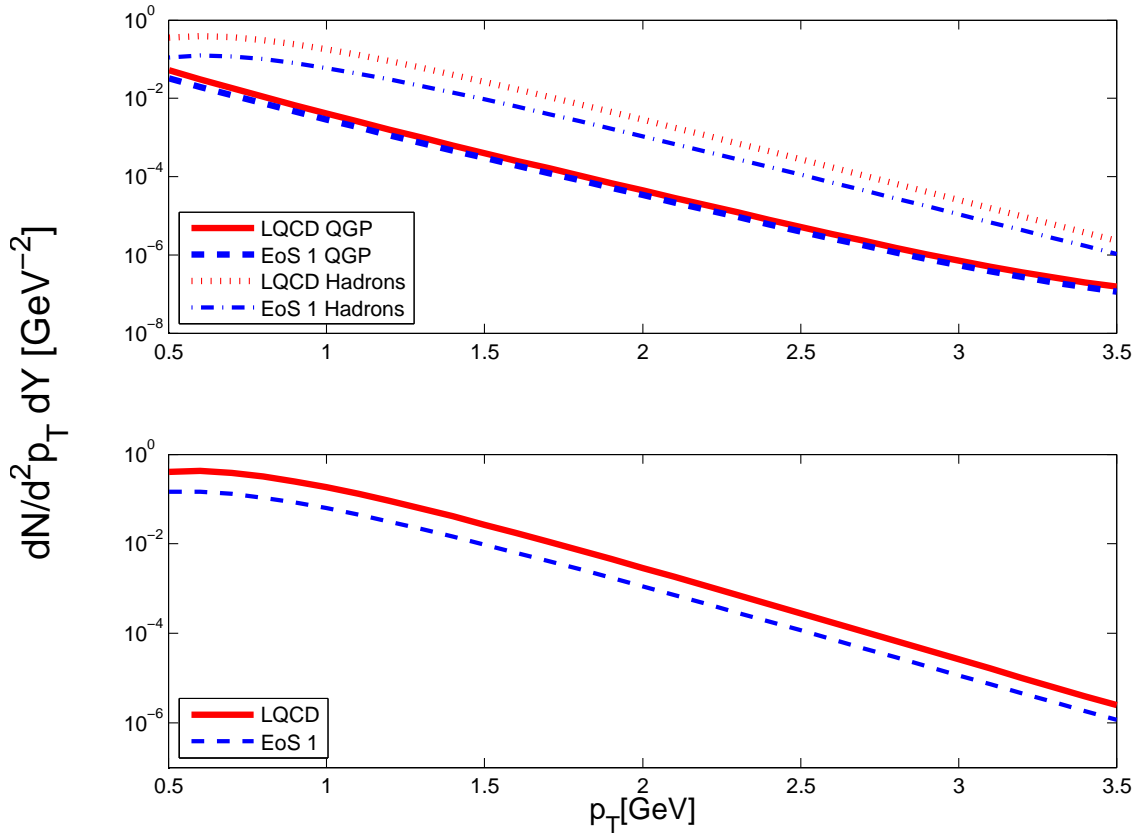}}
\vspace*{8pt}
\caption{(Color online) Thermal photon spectrum obtained from simulations using the EOS of Laine and Schr\"{o}der$^{43}$ (denoted by LQCD) and an EOS computed from a simple chiral--hydrodynamic model$^{23}$ (denoted by EOS 1). The upper panel shows the contributions corresponding to the QGP and hadronic phases, while the lower panel shows the total spectrum. The results correspond to a temperature--independent $\eta/s=0.11$ (Model A).}
\label{figap}
\end{figure}

To this purpose, we present results of photon spectra obtained from fluid dynamic simulations that use two different EOSs as input. Figure \ref{figap} shows the spectrum of thermal photons obtained using the EOS of Laine and Schr\"{o}der \cite{laine} (we will call it LQCD EOS in what follows) and the EOS shown in the lower panel of Figure \ref{eta}. The latter EOS corresponds to the linear sigma model and was obtained from the linearized Boltzmann equation in \cite{jpr-krein} (we will call it EOS 1), and is used here for illustrative purposes. We note that EOS 1 has been used in \cite{jpr-krein} to study the influence of chiral fields on charged hadron observables. 
The upper panel of Figure \ref{figap} shows the contribution of the QGP and hadronic phases to the total spectrum, which is shown in the lower panel. To keep things simple, we present the comparison for a temperature--independent $\eta/s = 0.11$ (corresponding to Model A).

It is seen that the main difference between the spectra obtained with the EOSs comes from the hadronic phase, with the photon spectrum being considerably smaller in the case of EOS 1. The difference between the spectra in both models is practically independent of $p_T$, except at very low values where this difference becomes slightly larger.
There are two reasons that make the spectrum obtained from the EOS 1 model smaller than the corresponding to the LQCD EOS. First, the hydrodynamic evolution is much faster in the EOS 1 model, essentially because the speed of sound is on average larger than the one corresponding to the LQCD EOS. This is illustrated by the fact that the  freeze-out temperature is reached in $\sim 6$ fm/c in the EOS 1 model and in $\sim 10$ fm/c in the  LQCD EOS, which represents a significant difference. The other reason is that the shear tensor $\Pi^{\mu\nu}$ is on average much smaller in the EOS 1 model, making the dissipative contribution to photon production smaller.
Therefore, the spectrum of thermal photons is sensitive to the EOS used in the fluid dynamic simulations in two ways, namely through viscous corrections to the spectrum and through the rate of expansion and cooling of the fireball. 

Comparing the results for different parametrizations for the temperature dependence of $\eta/s$ with those obtained using different EOSs, it is seen that the effect of the temperature dependence of $\eta/s$ on thermal photon spectra becomes appreciable only at $p_T \geq 1.5$ GeV, whereas the effect of using different EOSs is significant in the whole range of values of $p_T$ in the case of the hadronic contribution and very small for the QGP contribution. Moreover, it is also seen that when the difference between the results obtained using Models A, B and C is appreciable, the influence on the spectra of the temperature dependence of $\eta/s$ and of the EOS is comparable. 
This implies that, if one would attempt to extract $\eta/s$ by matching the thermal photon spectrum computed from fluid dynamic simulations to data, the uncertainty associated to the temperature dependence of the ratio and to the EOS used in the simulations would be comparable. 
Since neither the EOS nor the temperature dependence of $\eta/s$ of nuclear matter created in heavy ion collisions is precisely known, it would prove quite difficult to extract the temperature dependence of $\eta/s$ from measured photon spectra. Our results highlight the necessity of carrying out a combined study of different observables (photonic and hadronic) in order to obtain more accurate constraints on the temperature dependence of the shear viscosity--to--entropy--ratio of matter created in heavy ion collisions.

\section{Conclusions}
\label{conc}

In this paper we have computed the spectrum of thermal photons including dissipative corrections in the production rates, using three parametrizations for the temperature dependence of $\eta/s$ that enter as input in fluid dynamic simulations that employ a realistic Lattice QCD equation of state. We have also compared the influence of  the temperature dependence of $\eta/s$ on spectra to that of using different equations of state in the hydrodynamic simulations. 

Our results suggest  that some caution must be exercised when attempting to extract values of $\eta/s$ by fitting viscous fluid dynamic simulations to measured photon spectra.  By using simple parametrizations for $\eta/s$ as a function of temperature, we have shown that the temperature dependence of this ratio has a significant influence on the thermal photon spectrum, which increases with increasing transverse momentum, becoming significant for $p_T \geq 1.5$ GeV. Probably the most important result that we obtain is that the spectrum of thermal photons depends significanlty on the values of $\eta/s$ in the QGP phase but is quite insensitive to the values in the hadronic phase, a feature that deserves further study. Work is in progress along this line.

\section*{Acknowledgements}

We thank the anonymous referee for useful comments and suggestions that helped in improving the paper. 
We acknowledge Funda\c cao de Amparo a Pesquisa do Estado de S\~ao Paulo (Brazil) for financial support.

\end{document}